\documentclass[12pt,preprint]{aastex}

\shorttitle{The cm-wave continuum in LDN~1622}
\shortauthors{Casassus et al.}

\begin{document}

\title{Morphological analysis of the cm-wave continuum in the dark
cloud LDN~1622}

\author{S.~Casassus\altaffilmark{1}, G.~Cabrera, F.~F\"orster\altaffilmark{2}}
\affil{Departamento de Astronom\'{\i}a, Universidad de Chile,
Santiago, Casilla 36-D, Chile}
\author{T.J.~Pearson, A.C.S.~Readhead, C.~Dickinson}
\affil{Owens Valley Radio Observatory, California Institute
  of Technology, Pasadena, CA 91125}

\altaffiltext{1}{email address: \tt simon@das.uchile.cl}
\altaffiltext{2}{Denys Wilkinson building, Physics Department, Oxford
University, Keble Road, Oxford OX1 3RH}
\email{simon@das.uchile.cl}

\begin{abstract}

The spectral energy distribution of the dark cloud LDN~1622, as
measured by Finkbeiner using {\em WMAP} data, drops above 30~GHz and
is suggestive of a Boltzmann cutoff in grain rotation frequencies,
characteristic of spinning dust emission.

LDN~1622 is conspicuous in the 31~GHz image we obtained with the
Cosmic Background Imager, which is the first cm-wave resolved image of
a dark cloud.  The 31~GHz emission follows the emission traced by the
four {\em IRAS} bands.  The normalised cross-correlation of the 31~GHz
image with the {\em IRAS} images is higher by 6.6~$\sigma$ for the
12~$\mu$m and 25~$\mu$m bands than for the 60~$\mu$m and 100~$\mu$m
bands: $C_{12 + 25} = 0.76 \pm 0.02 $ and $C_{60 + 100} = 0.64 \pm
0.01$.

The mid-IR -- cm-wave correlation in LDN~1622 is evidence for very
small grain (VSG) or continuum emission at 26--36~GHz from a hot
molecular phase. In dark clouds and their photon-dominated regions
(PDRs) the 12~$\mu$m and 25~$\mu$m emission is attributed to
stochastic heating of the VSGs.  The mid-IR and cm-wave dust emissions
arise in a limb-brightened shell coincident with the PDR of LDN~1622,
where the incident UV radiation from the Ori~OB~1b association heats
and charges the grains, as required for spinning dust.


\end{abstract}

\keywords{radio continuum: ISM,
radiation mechanisms: general,
infrared: ISM,
ISM: dust,
ISM: clouds}

\section{Introduction}

An increasing amount of evidence supports the existence of a new
continuum emission mechanism in the diffuse interstellar medium (ISM)
at 10--30~GHz, other than free-free, synchrotron, or an hypothetical
Rayleigh-Jeans tail of cold dust grains\footnote{such traditional
grain emission is that due to thermal oscillations of the grain charge
distribution \citep[e.g.][]{dl99}}
\citep{lei97,deol99,fin99,deol02,lag03,ban03,fin04}. Examples of
excess emission at cm-wavelengths over known emission mechanisms have
been found in the spectral energy distributions (SEDs) of the dark
cloud LDN~1622 and the diffuse H\,{\sc ii} region LPH 201.7+1.6
\citep{fin02, fin04}, in the Helix planetary nebula \citep{cas04}, and
in another diffuse H\,{\sc ii} region in Perseus \citep{wat05}. At the
date of writing, the only morphological evidence for the existence of
a new emission mechanism at cm-wavelengths in a specific object is
provided by the Helix nebula. But a comparative analysis of the
cm-wave, mid- and far-IR continua in the Helix is hampered by strong
line contamination in the short wavelength {\em IRAS} maps. The Cosmic
Background Imager (CBI) observations of LDN~1622 provide an
opportunity of performing such morphological analysis.

As modelled by \citet{dl98a,dl98b} a possible candidate mechanism is
electric dipole emission from spinning very small grains (VSGs), or
`spinning dust'. The spectral energy distribution of the dark cloud
LDN~1622 \citep[Lynds Dark Nebula,][]{lyn62} is suggestive of spinning
dust: it rises over 5--9.75~GHz \citep{fin02}, following dipole
emission, and then drops above 30~GHz \citep{fin04}, as expected from
a Boltzmann cutoff in the grain rotation frequencies.

The dark cloud LDN~1622 lies within the Orion East molecular cloud
\citep{mad86}, at a distance of $\sim$120~pc \citep{wil05} and in the
foreground of the Orion B cloud. Its far-IR linear size is slightly
less than 1~pc.  It is a conspicuous CS$(2-1)$ and N$_2$H$^+$
``starless''\footnote{LDN~1622 does contain entries in the {\em IRAS}
Point Source Catalog, and probably hosts low-mass young stellar
objects, see Appendix~\ref{sec:midirps}} core \citep[with an H$_2$
density of $\sim 10^{3}-10^{4}$~cm$^{-3}$]{lee01}. LDN~1622 is devoid
of H\,{\sc ii} regions, aside from Barnard's Loop
\citep[e.g.][]{bou01}, a very diffuse H\,{\sc ii} region \citep[with
electron density of 2~cm$^{-3}$]{hei00} separated by $\sim$1~deg from
LDN~1622. No free-free emission is expected from LDN~1622, which is
indeed absent from the Parkes-MIT-NRAO survey at 5~GHz\footnote{Given
that the 1~$\sigma$ noise level in the PMN survey is
5~mJy~beam$^{-1}$, the free-free emission measure towards LDN~1622
must be less than 10~pc~cm$^{-6}$, which for a spherical nebula
10~arcmin in diameter implies electron densities of less than
10~cm$^{-3}$} \citep[hereafter PMN survey,][as presented in {\em
SkyView}, {\tt http://skyview.gsfc.nasa.gov}]{co93}.  Only the
H$\alpha$ corona of LDN~1622, outlining its photon-dominated region,
is marginally detected in the PMN survey.




Here we present the first cm-wave continuum image of a dark cloud, and
report morphological evidence that supports spinning dust as the
mechanism responsible for the anomalous foreground. We first describe
data acquisition (Section~\ref{sec:obs}), and image reconstruction
(Sec.~\ref{sec:syn}), and then discuss the effects of ground spill
over and give flux estimates (Sec.~\ref{sec:ground}). We analyse the
31~GHz data by comparison with the {\em IRAS} bands used as templates
for the emission by cool dust and by VSGs (or hot dust,
Sec.~\ref{sec:xcorr}), which leads us to infer a limb-brightened
morphology of LDN~1622 at 31~GHz.  The comparison with H$\alpha$ and
5~GHz templates shows that any free-free contribution at 31 GHz is
negligible, and that the 31~GHz emission is interior to the H$\alpha$
corona of the cloud (Sec.~\ref{sec:ha}).  We discuss the spectral
energy distribution of LDN~1622 (Sec.~\ref{sec:spec}), and finally
summarise our results (Sec.~\ref{sec:conc}).

\section{CBI observations} \label{sec:obs}

The CBI \citep{pad02} is a planar interferometer array with 13
antennas, each 0.9~m in diameter, mounted on a 6~m tracking platform,
which rotates in parallactic angle to provide uniform
$uv$-coverage. The CBI receivers operate in 10 frequency channels,
with 1~GHz bandwidth each, giving a total bandwidth of 26--36~GHz. It
is located in Llano de Chajnantor, Atacama, Chile.

\begin{figure}
\epsscale{.5}  \plotone{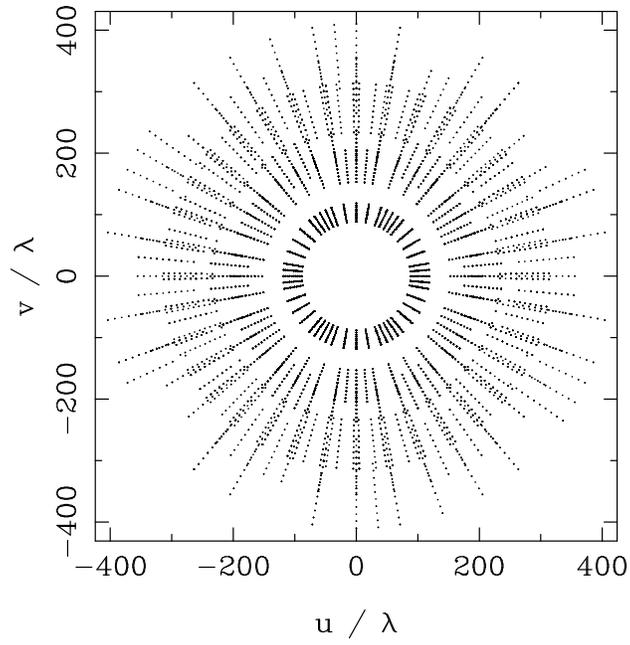}
\caption{$(u,v)$ coverage of the CBI in the compact configuration used
  for the observations of LDN~1622. \label{fig:uvcov}}
\end{figure}

We observed LDN~1622 (J2000 RA: 05:54:23.0, Dec: +01:46:54) on
03-Nov-2003, 02-Dec-2003 and 21-Nov-2004, for a total 10000~s.  The
compact configuration of the CBI interferometer results in the $(u,v)$
coverage shown on Fig.~\ref{fig:uvcov}, where it can be seen that
baseline length varies between 100~$\lambda$ and 400~$\lambda$,
corresponding to spatial scales of 34.4~arcmin and 8.6~arcmin,
respectively. Each receiver is equipped with phase shifters which
allow selecting its polarization mode. We set all receivers to L
polarization, so that the visibilities are sensitive to the
combination of Stokes parameters $I-V$. In what follows we assume that
Stokes $V$ (circular polarization) is negligible in LDN~1622. 
\clearpage
\clearpage
Approximate cancellation of ground and Moon contamination was obtained
by differencing with a reference field at the same declination but
offset in hour angle by the duration of the on-source integration. We
used an on-source integration time of 8~min, with a trailing reference
field. For phase calibration purposes we interspersed a 2~min
integration on J0607--085 between each 16~min cycle of differenced
observations.  J0607--085 was observed with identical telescope
settings as LDN~1622.

The data were reduced and edited using a special-purpose package
(CBICAL, developed by T.J. Pearson).  Flux calibration was performed
using either Saturn or Tau~A, whose fluxes are in turn calibrated
against Jupiter \citep[with a temperature of 146.6~K,][]{pag03}.  The
flux calibrator is also used as the reference for an initial phase
calibration. The phase calibration was subsequently refined by using
the calibrator interspersed between each cycle on LDN~1622. We applied
a phase shift to bring J0607--085 to the phase center. The magnitude
of the offsets by which we had to correct the position of J0607--085
varied between 15 and 40~arcsec.

In a final stage we combined all available visibilities of LDN~1622 to
produce two final datasets, with and without reference field
subtraction. Since the angular distance of LDN~1622 from the Moon was
larger than 80~deg for all three nights of observations, the
contamination on the shorter baselines is probably entirely due to
ground spill over.

\section{Image reconstruction} \label{sec:syn}

Image reconstruction is difficult for an object such as LDN~1622,
which extends to about half the CBI's primary beam of 45~arcmin FWHM,
and is surrounded by diffuse emission. Additionally the CBI's
synthesized beam obtained with natural weights, $\sim 8$~arcmin FWHM,
is about the size of the object, $\sim 10$~arcmin.  Thus in order to
perform a morphological analysis we need to extract a finer resolution
from the visibilities than that obtained from the restored images. The
maximum entropy method (MEM) fits model images to visibility data.
The MEM models can potentially recover details on finer angular scales
than the synthesized beam.  In this Section we present the results of
our reconstructions. The algorithm and model validation are described
in Appendices~\ref{sec:memalgo} and \ref{sec:modval}

\begin{figure}
\epsscale{1.}
\plotone{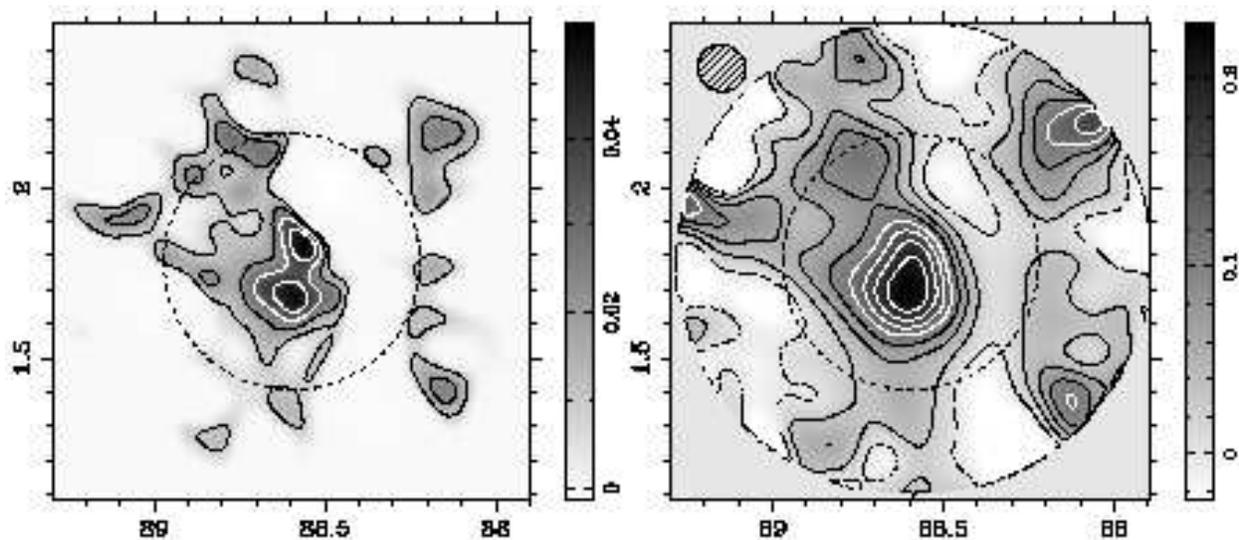}
\caption{Left: MEM model of the CBI data, specific intensity units are
  MJy~sr$^{-1}$, and contour levels are at $[0.010, 0.020, 0.031,
  0.042]$. Right: restored imaged obtained by convolving the MEM model
  with a gaussian PSF and adding the dirty map of the residual
  visibilities, specific intensity units are Jy~beam$^{-1}$. The
  contour levels are at $ [0, 0.029, 0.057, 0.086, 0.115, 0.144,
  0.172, 0.201] $.  Both the PSF (8.43~arcmin$\times$8.11~arcmin) and
  the natural-weight residual image were calculated with difmap. The
  half-power contour of the primary beam is shown as a dashed circle
  on both plots.
  \label{fig:cbi}}

\end{figure}

In Figure~\ref{fig:cbi}a we present a MEM model of our data.  The
noise of the restored image in Fig.~\ref{fig:cbi}b is close to that
expected from the instrumental noise.  The difmap package
\citep{she97} estimates a theoretical noise in the dirty map (using
natural weights) of 3.2~mJy/beam, which should give 3~$\sigma$
deviations of about 10~mJy~beam$^{-1}$ for an optimal
reconstruction. The dirty map of the residual visibilities, obtained
with difmap using natural weights, has a minimum of $-$11~mJy/beam
within the half-power contour of the primary beam, consistent with the
theoretical noise\footnote{the minimum value in the residual image is
in fact -24~mJy~beam$^{-1}$, at J2000 equatorial coordinates
(89.15~deg,+2.10~deg), which we identify as a 170~mJy point source
(PMN~J0604+0205) offset by 38~arcmin from the phase center in the
reference field. This negative point source in the restored image is
at $\sim$0.5~deg from the phase centre, and is outside the region of
interest}.
\clearpage
\clearpage

\begin{figure}
\epsscale{1.}
\plotone{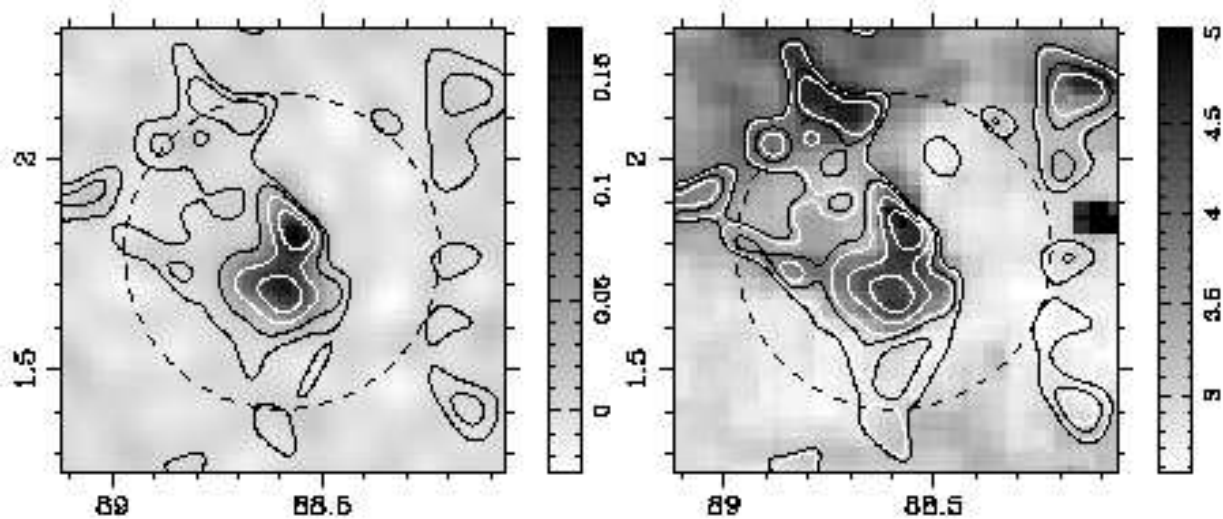}
\caption{left: Overlay of the MEM model (with the same contour levels
  as in Fig.~\ref{fig:cbi}) on a ``clean'' restoration of the CBI
  data, obtained with uniform weights in difmap. Units of the grey
  scale are Jy~beam$^{-1}$, with a $6.14\times5.8$~arcmin$^2$ beam
  (uniform weights). right: Overlay of the MEM model contours on the
  {\em IRAS}~12~$\mu$m map (with the same contours as in
  Fig.~\ref{fig:cbi}, and two extra levels in black, at $[0.008,
  0.0168]$). Note the MEM model traces 12~$\mu$m diffuse emission, but
  not the 12~$\mu$m point sources.
  \label{fig:clean}}
\end{figure}

We also obtained ``clean'' images with difmap, which qualitatively
confirm the MEM models. We show an overlay of the MEM model on a
``clean'' restoration in Fig.~\ref{fig:clean}a, obtained with the
difmap package and uniform weights. We anticipate from
Sec.~\ref{sec:xcorr} the good match between 31~GHz and 12~$\mu$m
emission to test which of the two reconstructions, whether MEM or
``clean'', extracts the most of the data.  Fig.~\ref{fig:clean}b also
shows an overlay of the 31~GHz contours on the {\em IRAS} 12~$\mu$m
map in grey scale. The {\em IRAS} 12~$\mu$m image is from the IRAS Sky
Survey Atlas \citep{whe91}, as obtained in {\em SkyView}.  It can be
appreciated by inspection of Fig.~\ref{fig:clean} that the MEM model
recovers low-level details that are absent in the ``clean'' image,
such as the 12~$\mu$m emission peaks at $(88.8,+2.1)$ and
$(88.1,+2.1)$. There are two features in the MEM model that do not
seem to have a 12~$\mu$m counterpart. One is a low-level contour at
$(88.8,+1.3)$, which turns out to be the location of the brightest
radio point source in the field (see Sec.~\ref{sec:ha} and
Fig.~\ref{fig:shassa}). The other is a 31~GHz peak at $(88.1,+1.4)$,
which matches an H$\alpha$ feature at the outskirts of Barnard's Loop
(see Fig.~\ref{fig:shassa}). We will use MEM in what follows because
it provides reconstructions that do not depend on user-defined
``clean'' boxes, and because it allows extracting details on fine
angular scales while preserving the sensitivity of the
dataset\footnote{the MEM algorithm implemented here does not apply any
gridding, so that the visibilities are assigned their statistical
weight only. To reach finer angular resolutions, ``clean''
reconstructions downweight low spatial frequencies, thereby loosing
sensitivity.}.

\clearpage
\clearpage

%

The CBI image can be compared with that available in the first {\em
WMAP} data release. Barnard's loop is the most conspicuous feature in
the {\em WMAP} Ka-band image of the region. But it is apparent that
the CBI data on LDN~1622 are much more sensitive, and allow resolving
the dark cloud. The CBI image is thus the first at cm-wavelength of a
dark nebula, i.e. a cold dust cloud identified by visible-light
stellar counts.

%
%
%


\section{Ground  contamination and average properties of the
dataset} \label{sec:ground}

\begin{figure}
\epsscale{1.}
\plottwo{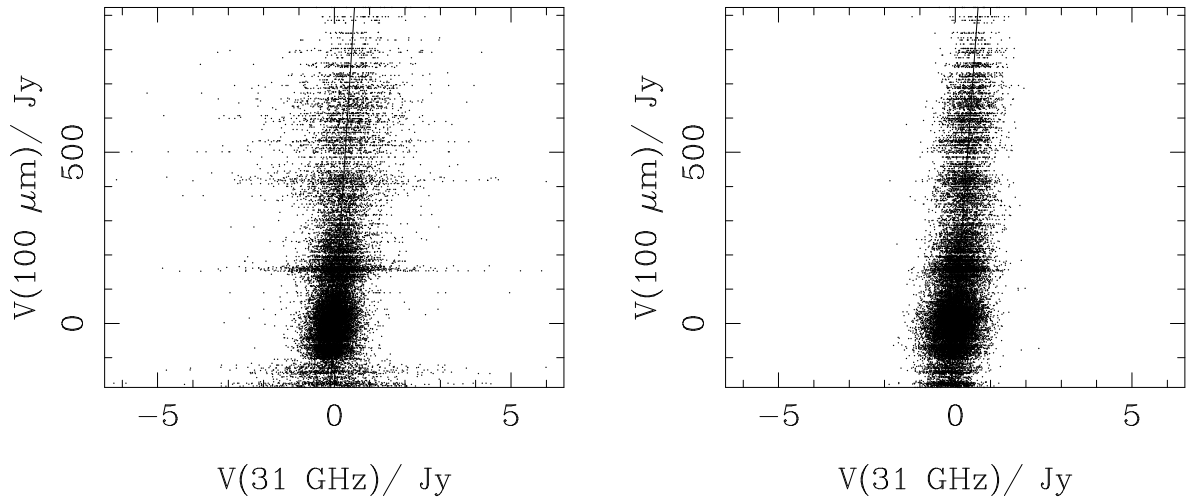}{f4b.eps}
\caption{31~GHz-100$\mu$m visibility correlations over the full range
  of $uv$-radii, for the non-differenced dataset (left) and for the
  differenced dataset (right). We plot both the real and imaginary
  parts. \label{fig:xcorr}}
\end{figure}

In order to cross-correlate the CBI data with the comparison templates
we compute template visibilities, obtained by a simulation of CBI
observations on the template images (``CBI-simulated visibilities''
herafter, see Appendix~\ref{sec:memalgo} and \ref{sec:modval}). The
31~GHz-100~$\mu$m visibility plot on Fig.~\ref{fig:xcorr}a for the
undifferenced dataset allows assessing the level of ground and Moon
contamination in the shorter baselines. The enhanced scatter above
100~$\mu$m visibilities of $V_{100\mu\mathrm{m}}=300~$Jy and at
$V_{100\mu\mathrm{m}}=\pm~150~$Jy is suppressed in the differenced
dataset shown on Fig.~\ref{fig:xcorr}b.
\clearpage
\clearpage

\begin{figure}
\epsscale{.5}
\plotone{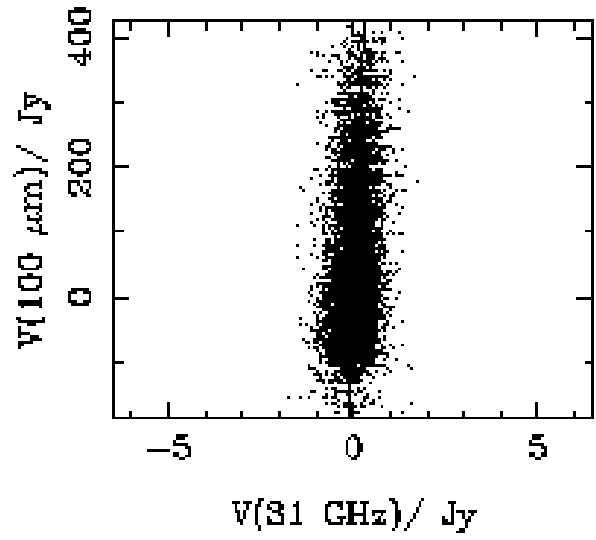}
\caption{31~GHz-100$\mu$m visibility correlation for the
  non-differenced dataset and $uv$-radii $ k > 120$.
  \label{fig:xcorr_uvrange}}
\end{figure}

The enhanced scatter due to ground or Moon contamination in the
shorter baselines correspond to where the real parts of
$V_{100\mu\mathrm{m}}$ reach about 300~Jy and where the imaginary
parts of $V_{100\mu\mathrm{m}}$ reach $\pm~150$~Jy. Restricting to
baselines above 120~$\lambda$ retains visibilities devoid of ground
contamination, as shown on Fig.~\ref{fig:xcorr_uvrange}.
\clearpage
\clearpage

Typical ISM power spectra are decreasing power-laws\citep{gau92,
wri98, elm02}, the ensemble-averaged modulus of the visibility is thus
a monotonic function of $uv$-radius. This is also true for LDN~1622,
for the case of the 100~$\mu$m CBI-simulated visibilities: the
azimuthally averaged power-spectrum is monotonic.  We cannot recover a
power spectrum for the CBI data by simple averaging because the signal
is affected by noise, so that the derived spectrum is artificially
flat.

The 31~GHz flux density measured on the restored image within a
circular aperture with 45~arcmin diameter centered on LDN~1622 is
1.41$\pm$0.03~Jy.  We caution that the CBI images are heavily affected
by flux loss for emission on 45~arcmin scales: because of incomplete
sampling in the $uv$ plane the reconstructed images have missing
spatial frequencies, and part of the extended nebular emission is
lost. We can infer a flux density corrected for flux loss, of $2.90
\pm 0.04$~Jy, by referring to a template map, for which we use {\em
IRAS}~100$\mu$m. This flux density is estimated by extracting the flux
density in the template map within a 45~arcmin aperture, and scaling
by the CBI-{\em IRAS}~100$\mu$m correlation slope given in
Table~\ref{table:corr} for the differenced dataset.



Template maps which follow closely the 31~GHz emission also allow a
cross-check on the pointing accuracy of the CBI. We vary an $( \alpha
, \delta )$ shift on the coordinates of the reference pixel of the
template maps to minimize $\chi^2 = \| V_i(31~\mathrm{GHz}) -
V_i(IRAS) \| ^2/\sigma_i^2 $, where the uncertainties $\sigma_i$ only
contain the CBI noise. The optimal shifts we find for each of the {\em
IRAS} maps are (in arcmin): ( $ 0.66\pm 0.07$, $ 0.24\pm 0.08$ ), ( $
0.60\pm 0.08$, $ 0.14\pm 0.09$ ), ( $ 0.79\pm 0.08$, $-0.31\pm 0.09$
), ( $ 0.08\pm 0.08$, $-0.89\pm 0.09$ ) for the 12~$\mu$m, 25~$\mu$m,
60~$\mu$m, and 100~$\mu$m maps, respectively. Thus no particular trend
is found, although the average values of the shifts is
$(+0.5,-0.3)$~arcmin and significantly different from zero. But in
what follows we ignore a possible residual error in telescope pointing
because the overlays of the 31~GHz and far-IR images in
Fig.~\ref{fig:templates} show a good match, and would not improve by
shifting on 0.5~arcmin scales.

\section{Comparison with mid- and far-IR templates.} \label{sec:xcorr}
 
If dust is reponsible for the 31~GHz emission in LDN~1622 then a tight
relationship is expected with the mid- and far-IR emission. Here we
investigate the consequences of assuming that the emission traced by
the CBI scales linearly with the four {\em IRAS} maps.

The infrared emission from dust is discussed in details by, e.g,
\citet{des90,dra01,li01}. The {\em IRAS}~100$\mu$m band traces
emission from large grains, with sizes greater than 0.01~$\mu$m.  The
large grains are in equilibrium with the interstellar radiation field,
with a temperature of order 10--20~K depending on environment.
Continuum emission at shorter wavelengths is due to hot dust, at
$\sim$100~K, which is too hot to be maintained in equilibrium with the
interstellar UV field. Mid-IR emission from classical hot dust is not
expected because of the absence of a strong UV source within LDN~1622,
in contrast with compact H\,{\sc ii} regions or planetary
nebulae. Thus stochastic heating of VSGs dominates the dust emission
in the {\em IRAS} 12$\mu$m and 25~$\mu$m bands. The heat capacity of a
VSG is so small that the absorption of a single UV photon increases
the particle temperature high enough for it to emit at $<60\mu$m.

Thus, by examing the degree of correlation with the 4 {\em IRAS}
bands, we hope to determine which type of grain, whether the large
grains or the VSGs, are responsible for the 31~GHz emission. We
caution that from the {\em IRAS} photometry alone we cannot
differentiate a 31~GHz link to the VSGs from a link to a hot molecular
phase that shines in the H$_2$ lines. After all the VSGs can also be
regarded as large molecules, such as polycyclic aromatic hydrocarbons
(PAHs). Another important source of flux in the {\em IRAS}~12$\mu$m
and 25$\mu$m bands are the H$_2$ ro-vibrational lines, such as
H$_2$(0--0)S(2)~12.3$\mu$m, and H$_2$(0--0)S(0)~28.2$\mu$m. The H$_2$
line fluxes integrated over the {\em IRAS} band passes could account
for part of the mid- and far-IR morphological differences \citep[as
could be the case in PDRs, with conspicuous H$_2$ lines,][]{vand04}.

The 31~GHz MEM contours can be compared by inspection with the raw
{\em IRAS} images, as extracted from {\em SkyView}.  The diffuse
emission in the mid-IR images is closer to the 31~GHz contours than
the far-IR images. In this section we quantify this qualitative
result, and show it is not affected by noise or missing spatial
frequencies at 31~GHz.



\subsection{Visibility cross-correlations}

Is the cm-wave -- mid-IR correlation detectable directly in the
visibility data? The cross-correlations may be different in the image
plane and in the $uv$ plane because of two reasons. One is the
contribution of point sources at 12~$\mu$m, which are absent at
31~GHz. The fainter point sources at 12~$\mu$m may be numerous and act
as diffuse emission\footnote{see for instance the ISO 6.7~$\mu$m image
of LDN~1622 in Fig.~\ref{fig:isocam}}. The subtraction of the
brightest point sources may not be accurate enough to retain genuinely
diffuse emission at 12~$\mu$m.  Another difficulty with a $uv$-plane
analysis of the diffuse emission are the uncertainties in the CBI
primary beam \citep[][their Fig.~1]{pea03}. Variations between
antennas introduce uncertainties beyond about 40~arcmin from the phase
center. LDN~1622 is surrounded by diffuse emission, such as that
traced by {\em IRAS}. The outskirts of Barnard's Loop are within
35~arcmin from the phase center, and it peaks at about 60~arcmin. Also
in the neighborhood of LDN~1622 is the reflection nebula NGC~2067,
which at 1.9-2.7~deg from the phase center falls on a sidelobe of the
primary beam at 2.2~deg.  Barnard's Loop or NGC~2067 are bound to
enter the side lobes and low level wings of the primary beam, where
the uncertainties in the primary beam model used in CBI-simulated
visibilities become important.

We linearly correlate the CBI visibilities with the CBI-simulated
visibilities on the four {\em IRAS} bands, one template at a time, and
after processing as described in
Appendix~\ref{sec:modval}. Table~\ref{table:corr} lists reduced-$\chi^2$,
linear correlation coefficients \citep[as defined in][]{bev92},
correlation slopes and uncertainties. In Table~\ref{table:corr} we
also consider the non-differenced CBI dataset, because its signal to
noise ratio (S/N) improves by a factor $\sqrt{2}$. We restrict the
analysis of the non-differenced dataset to $uv$-radii above
120~$\lambda$. Such a baseline range allows minimizing ground spill
over or Moon contamination. Another reason for restricting baseline
lengths above a minimum is that a constant background in the template
maps affects the simulated visibilities for the shortest baselines.
We minimize this effect, which is due to the restricted sky domain
available to compute the simulated visibilities, by clipping the
templates so that their minimum intensity value is zero.


\begin{deluxetable}{lllrr}
\tabletypesize{\small} \tablecaption{Linear correlation results
\label{table:corr}} \tablewidth{0pt} \tablehead{ & \colhead{12~$\mu$m}
& \colhead{25~$\mu$m} & \colhead{60~$\mu$m} & \colhead{100~$\mu$m} }
\startdata 
\multicolumn{5}{c}{A: differenced dataset,  $f = 25724$} \\
$\chi^2/f$: & 1.07 & 1.06 & 1.06 & 1.06 \\ 
 $r$: & 0.395  & 0.399 & 0.394 & 0.395  \\ 
 $a$: & $24.86\pm 0.30$ & $14.86\pm 0.18$ & $ 2.64\pm 0.03$
& $ 0.70\pm 0.01$     \\ \hline 
\multicolumn{5}{c}{B: non-differenced dataset, $k> 120$, $f = 16016$} \\
$\chi^2/f$: & 1.74   & 1.74 &  1.74 &  1.75  \\ 
 $r$: & 0.243  & 0.241  & 0.246  & 0.242  \\ 
 $a$: & $20.09\pm 0.46$ & $13.09\pm 0.30$ & $ 2.60\pm 0.06$
& $ 0.68\pm 0.02$   \\  \enddata 
\tablecomments{$f$ is the number of degrees of freedom (which is the
  number of observed visibilities above a $uv$-radius $k$, minus
  one free-parameter), $r$ is the linear  correlation coefficient, and
  $a$ is the conversion factor between the various templates and the
  31~GHz visibilities, such that $V(31~\mathrm{GHz})=10^{-3} ~a
  ~V(\mathrm{IR})$.}
\end{deluxetable}

But from the visibility correlations alone we cannot ascertain which
{\em IRAS} map correlates best with the 31~GHz data. The significance
of the results is difficult to assess because the noise in the
comparison maps is not known accurately (and is neglected in this
analysis), and the confidence level associated to the $\chi^2$
distribution with $\nu \sim 10\,000$ degrees of freedom is extremely
sharp at $\chi^2/\nu \approx 1$.

\subsection{Image plane cross-correlations}

A drawback of analysing the visibility data directly in the
Fourier-plane is that the effect of the point sources is difficult to
isolate, especially at shorter IR wavelengths, where point sources are
more frequent. Here we compare the CBI data and the {\em IRAS}
templates in the image plane, based on our MEM modelling.

\subsubsection{Qualitative comparison}

\begin{figure}
\epsscale{1.}
\plotone{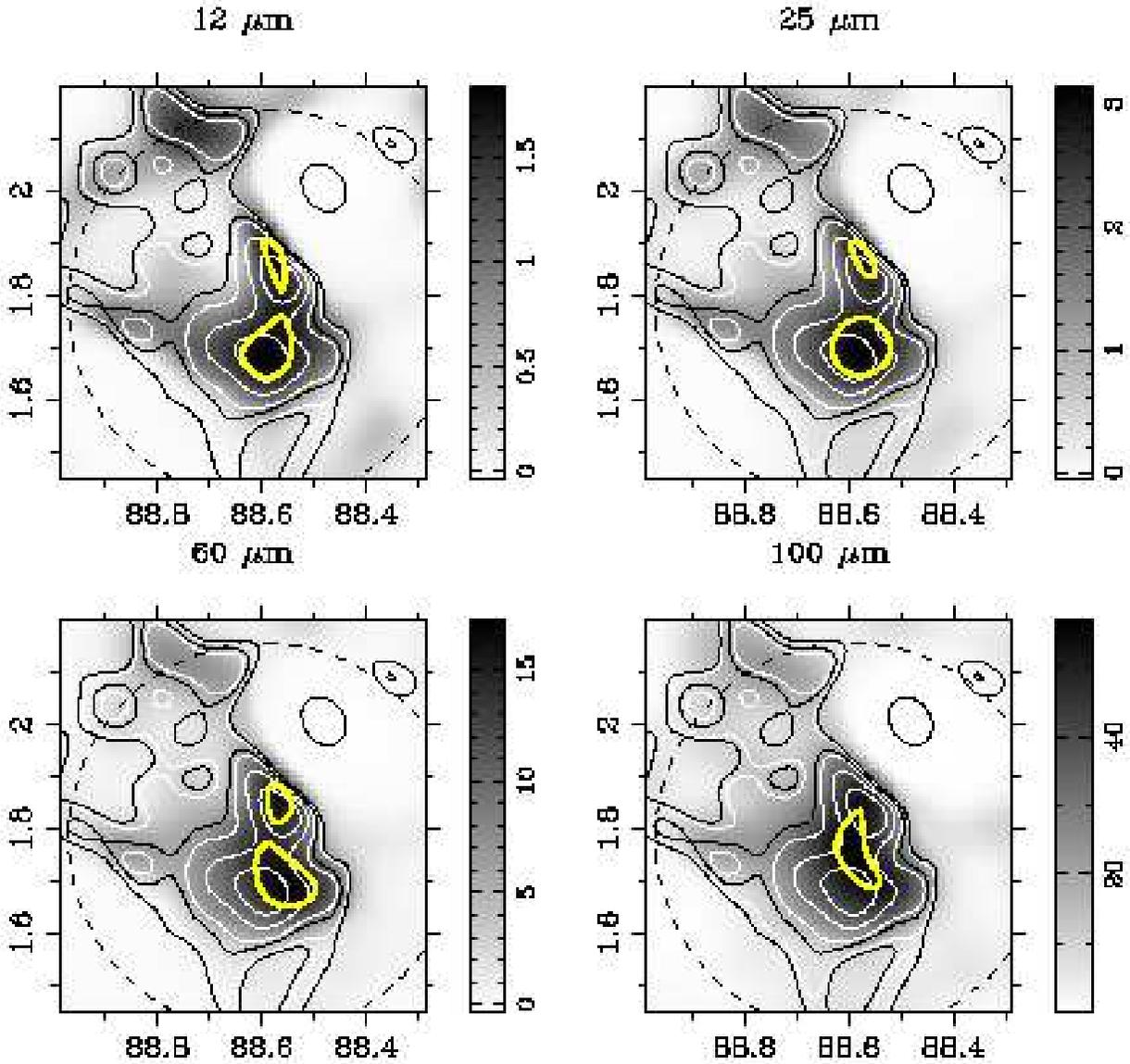}
\caption{All intensity units are MJy~sr$^{-1}$, the grey scales
  correspond to the MEM model for the {\em IRAS} templates, and the
  thin contours follow the CBI 31~GHz MEM model, as in
  Fig.~\ref{fig:clean}. The thick contours follow the {\em IRAS}
  bands, at a fraction of the peak intensity: 89\% for {\em
  IRAS}~12$\mu$m, 85\% for {\em IRAS}~25$\mu$m and 60$\mu$m, and 95\%
  for {\em IRAS}~100$\mu$m. \label{fig:templates}}
\end{figure}

Inspection of Fig.~\ref{fig:templates} reveals that the 12~$\mu$m and
25~$\mu$m MEM maps are the most similar to the 31~GHz MEM model.  In
Fig.~\ref{fig:templates} the {\em IRAS} maps are the same as in
Fig.~\ref{fig:irasmem}, i.e. they are reconstructed from simulated CBI
visibilities, following the algorithm described in
Appendix~\ref{sec:modval}.  Thus the mid-IR -- cm-wave correlation is
not the effect of missing spatial frequencies in the 31~GHz visibility
data.

The 100~$\mu$m emission is concentrated in a single maximum, while the
31~GHz, 12~$\mu$m and 25~$\mu$m images show two peaks near the phase
center, at the origin of coordinates in Fig.~\ref{fig:templates},
which we refer to as the northern and southern peaks. The 60~$\mu$m
image is also double-peaked, but the southern peak is offset relative
to the 31~GHz southern peak.

The 31~GHz morphology of LDN~1622 is remarkably similar to that in the
12~$\mu$m {\em IRAS} band. However there is an interesting feature at
12~$\mu$m which is absent at 31~GHz. The northen peaks at 31~GHz and
12~$\mu$m are slightly offset, while the southern peaks are exactly
coincident. We explain the shift in the position of the northen peak
as being due to a young stellar object (YSO), namely L1622-10, whose
emission contributes at 12~$\mu$m but not at 31~GHz. Thus the 31~GHz
emission is genuinely diffuse, while 12~$\mu$m includes photospheric
emission, or unresolved very hot dust. The point-source flux for
L1622-10, as listed in the {\em IRAS} Point Source Catalog, is
subtracted from the processed {\em IRAS}~12$\mu$m image used as
comparison template (see Appendix~\ref{sec:modval}). But the YSO is
still present in the processed image, even after subtraction, as can
be inferred by comparing the raw {\em IRAS}~12$\mu$m image in
Fig.~\ref{fig:clean} and the processed image in
Fig.~\ref{fig:templates}. The imperfect subtraction of L1622-10 is
probably due to an inaccurate catalog flux, perhaps due to the
uncertainties inherent in deriving a flux density for a point source
on top of a compact source, such as the northern peak.  A discussion
on the properties of this YSO is given in Appendix~\ref{sec:midirps}.

\clearpage
\clearpage

\subsubsection{Statistics of the 31~GHz and IR templates correlations}




In order to quantify the similarities that meet the eye when comparing
the 31~GHz and the {\em IRAS} templates, we compute the normalised
cross-correlation $C$ of the 31~GHz MEM images with the {\em IRAS}
models, one at a time:
\begin{equation}
C = \sum_i I_i(31~\mathrm{GHz}) I_i(IRAS) / \sum_i
I_i(31~\mathrm{GHz})^2,
\end{equation}
where the sums extend over all pixels in the model images. We can
estimate the significance of the results by calculating the scatter of
the cross-correlation $C$ for each of 90 different realisations of
noise on the template visibilities (see Appendix~\ref{sec:modval}). 

We attempted to assign a $\chi^2$ value to the comparison between the
CBI and the {\em IRAS} models. But the pixels in the model images are
correlated, and the covariance matrix is prohibitively large, with
$\sim~170^4$ elements, in the case of $170^2$ free-parameters per MEM
model. The tests we ran to estimate the covariance matrix from the
simulations described in Sec.~\ref{sec:syn} showed we need many more
than only 90 different noise realisations. We reached a suitable
accuracy on the covariance matrix only in the useless case of a model
image with $\sim$10 free parameters.



\begin{figure}
\epsscale{.5}
\plotone{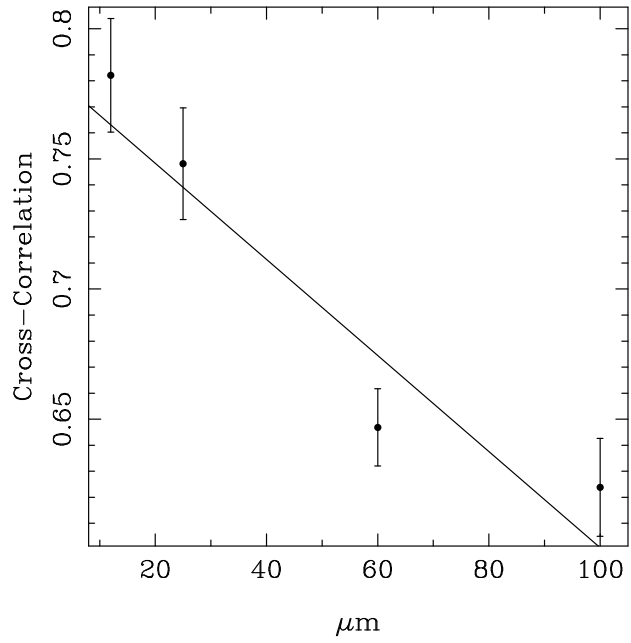}
\caption{The cross-correlation of the 31~GHz and {\em IRAS} images
  ($y-$axis) as a function of wavelength ($x$-axis). 
\label{fig:skyxcorr}}
\end{figure}

Table~\ref{table:skyxcorr} lists the cross-correlation results, which
are also summarised in Fig.~\ref{fig:skyxcorr}. The weighted average
of the 60 and 100~$\mu$m cross-correlations is worse than that of the
12~$\mu$m and 25~$\mu$m cross-correlations by 6.6~$\sigma$.  The solid
line on Fig.~\ref{fig:skyxcorr} has a slope of
$-1.84~10^{-3}\pm2.9~10^{-4}$, and is thus different from zero at
6.3~$\sigma$.  To test the hypothesis that $C$ is independent of {\em
IRAS} band we calculate $\chi^2 = \sum_{j=1}^{4} (C_j - \langle C
\rangle)^2 / \sigma_j^2$ with the data from
Table~\ref{table:skyxcorr}, where $\langle C \rangle$ is the weighted
average of the cross-correlations. Reduced $\chi^2$ is 7.7 for 3
degrees of freedom, which discards a constant value of the
cross-correlation as a function of {\em IRAS} wavelength.

We also carried out the same simulations but with a different entropy
term, $S_b$ (described in Appendix~\ref{sec:memalgo}), obtaining the
same results at lower significance.  In this case the mid-IR -- far-IR
difference is 3.7~$\sigma$, or 4.4~$\sigma$ after subtraction of the
YSO L1622-10.

\begin{deluxetable}{lrrrr}
\tabletypesize{\small} \tablecaption{Results from the
  cross-correlations in the image plane. 
\label{table:skyxcorr}} \tablewidth{0pt} \tablehead{ & \colhead{12~$\mu$m}
& \colhead{25~$\mu$m} & \colhead{60~$\mu$m} & \colhead{100~$\mu$m} }
\startdata 
$C$:  & 0.782 $\pm$ 0.022 &  0.748  $\pm$  0.021 & 0.647  $\pm$  0.015 & 0.624  $\pm$ 0.019  \\ \enddata
\tablecomments{uncertainties are 1~$\sigma$}
\end{deluxetable}

\clearpage
\clearpage

\subsection{Interpretation}

The comparison between the CBI image and the four {\em IRAS} bands
allows us to conclude that the CBI emission is best represented by
{\em IRAS}~12$\mu$m.  The morphology of the mid-IR {\em IRAS} maps is
suggestive of limb-brightening of a VSG-emitting shell coincident with
LDN~1622's PDR, as required by UV excitation of the VSGs.

LDN~1622 is a rather peculiar cloud in that its mid-IR emission is
limb-brightened. By contrast, LDN~1591 reaches higher 100~$\mu$m
intensities than LDN~1622, and yet is not detected by
\citet{fin02}. The facts that LDN~1591 is not limb-brightened and that
the 26--36~GHz emissivity is enhanced in LDN~1622 lead us to propose
that the 26--36~GHz emission stems from the photon-dominated region,
with abundant UV radiation. This scenario is consistent with
``spinning dust'', or electric dipole radiation from spinning VSGs
exposed to the incident UV radiation and charged by the photo-electric
effect. A possible test for this interpretation may derive from the
analysis of the cm-wave morphology of other limb-brightened clouds,
such as DC300-17 in Chamaeleon \citep{laur89}, which like LDN~1622
also harbors low-mass YSOs.

\begin{figure}
\epsscale{.5}
\plotone{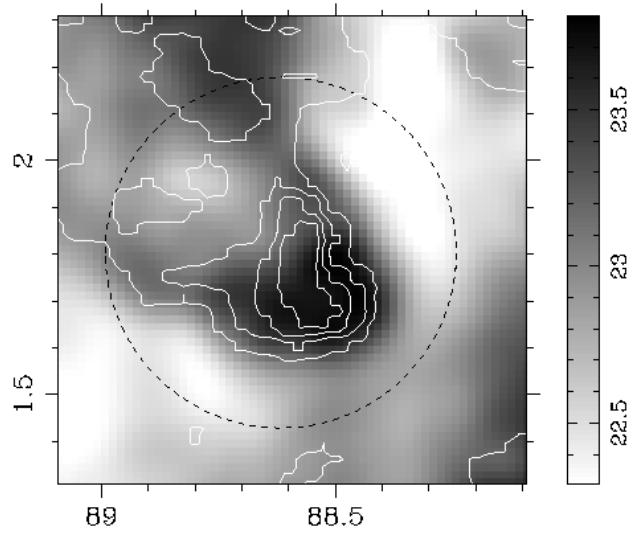}
\caption{ {\em IRAS}~60~$\mu$m contours overlaid on the dust
  temperature (in K) inferred from the {\em IRAS}~60 and 100~$\mu$m
  maps.
\label{fig:tdust}}
\end{figure}

Could ``magnetic dipole emission'' from large grains, proposed by
\citet{dl99}, also account for the mid-IR -- cm-wave correlation? The
fact that the 100~$\mu$m emission does not trace the 31~GHz
double-peaked morphology suggests large grains, with a modified
black-body spectrum, do not contribute at 31~GHz. An increased 31~GHz
emissivity through a temperature enhancement in the PDR of LDN~1622
would have a concomitant limb-brightened morphology at 100~$\mu$m.  We
infer a classical dust temperature map for LDN~1622, shown on
Fig.~\ref{fig:tdust}, from the {\em IRAS} 60/100~$\mu$m color map. We
adopted a $\nu^{2}$ emissivity law, and degraded the two maps to a
common resolution guided by the point sources in the field. We solve
for the dust temperature using the Brent method \citep{pre96}, pixel
by pixel. It is apparent that the large grain temperature is fairly
constant across LDN~1622, and does not follow the 31~GHz contours.

\clearpage
\clearpage

It may be argued that the {\em IRAS}~12~$\mu$m band is not well suited
to trace VSGs because it is often contaminated by ionic line emission,
for instance by [Ne\,{\sc ii}]~12.8$\mu$m, which could arise in the
PDR at the surface of LDN~1622.  But the similarity of the 12~$\mu$m
and 25~$\mu$m maps argue against significant line contamination. It
would be very contrived to have just the right contribution of flux
from lines in both bands (although the H$_2$ lines could still
contribute to both bands in similar proportions).


%
%

\section{Comparison with H$\alpha$ and 5~GHz templates} \label{sec:ha}

\begin{figure}
\epsscale{1.}
\plottwo{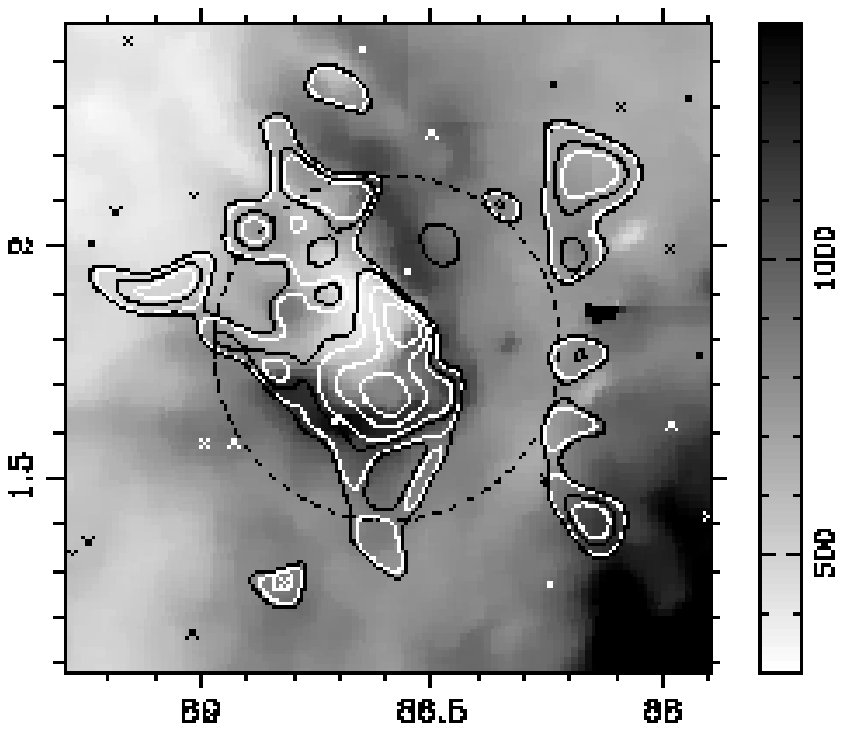}{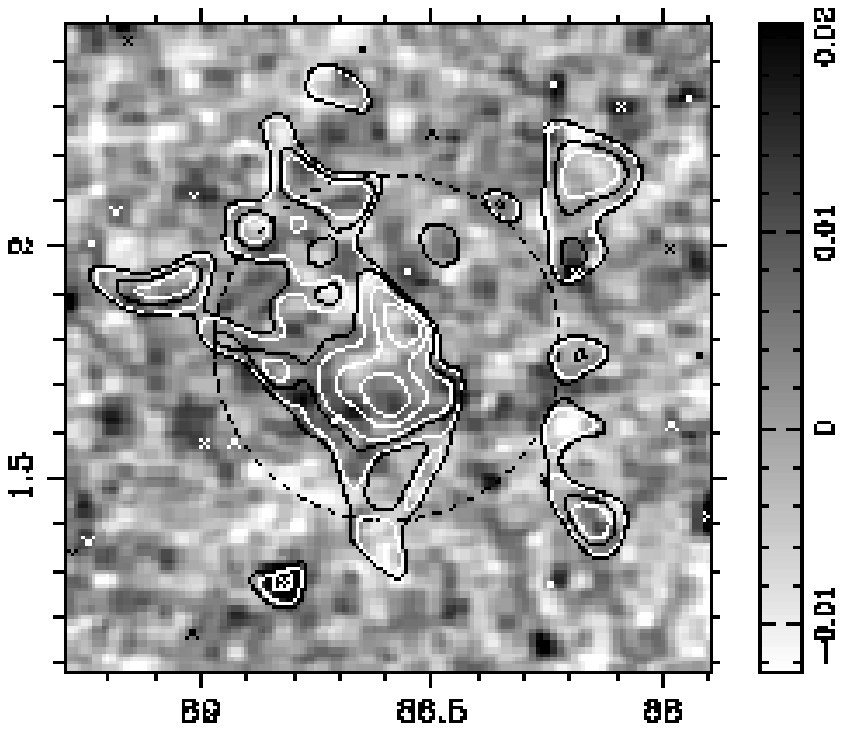}
\caption{Left: 31~GHz MEM model in contours overlaid on the H$\alpha$ map
  in gray scale, with intensity units in deci-Rayleighs. Right: 31~GHz
  MEM model in contours overlaid on the PMN 5~GHz map, in
  Jy~beam$^{-1}$. Crosses on both plots are the NVSS point sources,
  and the circle is the only point source from the North 6\,cm
  database.   
  \label{fig:shassa}}
\end{figure}

The surface of LDN~1622 is exposed to the interstellar UV field. Such
ionised corona of LDN~1622 is conspicuous in the SHASSA image
\citep{gau01}, shown on Fig.~\ref{fig:shassa}. The V-shaped H$\alpha$
corona points towards the Orion~OB~1b association \citep[][their
Fig.~1]{wil05}. We note LDN~1622 corresponds to a minimum in H$\alpha$
brightness, therefore it is a foreground object obscuring the diffuse
H$\alpha$ from the Orion-Eridanus bubble, consistent with the short
distance of \citet{wil05}.


That the 31~GHz emission is not free-free is apparent from
Fig.~\ref{fig:shassa}, where H$\alpha$ seems to anti-correlate with
the radio continuum, although we did not attempt to correct the
H$\alpha$ map for extinction. The H$\alpha$ and free-free emission
both trace electron-ion collisions, so that if the electron
temperature is constant, then the unreddenned H$\alpha$ intensities
are proportional to the radio-continuum specific intensities. But the
only correspondence between 31~GHz and H$\alpha$ is at (88.15,+1.4),
and stems from the outskirts of Barnard's Loop. There is no
counterpart of H$\alpha$ emission inside the CBI primary beam.

From the comparison with the PMN survery in Fig.~\ref{fig:shassa} we
further confirm free-free emission is negligible at 31~GHz.  There is
a hint of a radio counterpart of the H$\alpha$ corona, but no 5~GHz
emission coexists with the 31~GHz emission. 

We can further test the bremsstrahlung hypothesis for the 31~GHz
emission by extrapolating the observed intensity levels to 5~GHz with
a spectral index of $\alpha = -0.1$, in the optically thin
approximation.  The restored CBI image on Fig.~\ref{fig:cbi}b reaches
peak intensities of 0.22~Jy~beam$^{-1}$. Since the PMN beam is
3.7~arcmin FWHM\footnote{The PMN survey is published in Jy~beam$^{-1}$
units, but its resolution depends on whether the data were acquired
with the Green Bank 300-foot dish or with the Parkes 140-foot dish. We
calibrated the PMN survey with the 17 brightest point sources in a
6~deg field centered on LDN~1622, using as reference the fluxes listed
by \citet[][North 6cm catalog, also based on the PMN
survey]{bec91}. We fitted elliptical gaussians to each point source to
extract fluxes, and obtained that the beam solid angle used in the
intensity units must correspond to a 3.7~arcmin FWHM PSF to reproduce
the catalog fluxes. The average FWHM of the elliptical gaussians is
$3.78 \pm 0.56$~arcmin, coincident with the chosen intensity units.},
peak 5~GHz intensities should range from 264~mJy~beam$^{-1}$ for an
unresolved source, to 52~mJy~beam$^{-1}$ for a uniformly extended
source. The root-mean-square (rms) noise in the PMN image is $\sigma =
5.8~$mJy~beam$^{-1}$.  The absence of the 31~GHz features from the PMN
image therefore allows us to rule out free-free emission at
9~$\sigma$.

In order to assess possible contamination at 31~GHz by background
sources, we have overplotted on Fig.~\ref{fig:shassa} the entries from
the NVSS catalog \citep{con98} with flux densities greater than
10~mJy, as well as the entries from the North 6~cm database
\citep{bec91}. Only one source may be present at 31~GHz. This is PMN
J0555+0116, or NVSS J055516+011622 (J2000 RA: 05:55:16.62, Dec:
+01:16:22.9), which is the source at (88.72, 1.28) in
Fig.~\ref{fig:shassa}b, and well outside the dark cloud and the CBI
primary beam.


\citet{co93} explain that low spatial frequencies, on scales larger
than 30~arcmin in declination, are filtered-out from the PMN
survey. But the CBI-PMN comparison is not affected by this filter. The
PMN filter corresponds to the very largest angular scales observed by
the CBI, and LDN~1622 is a compact object of order 10~arcmin in
diameter. The outskirts of Barnard's Loop, picked up in the MEM model
at $(88.1,+1.4)$, is probably filtered-out in PMN.

\clearpage
\clearpage


\section{Spectral properties} \label{sec:spec}


\subsection{Low-frequency spectral index} \label{sec:lowfreqindex}

For a comparison with \citet{fin02} we must consider the consequences
of differencing on their chopped observations, which filters-out low
spatial frequencies. The flux densities in \citet{fin02} are referred
to the {\em IRAS}~100~$\mu$m map by linear cross-correlation. But in
general the radio and IR emissions are bound to have different
power-spectra in the ISM at large. Thus in general the slopes of the
straight line fits between radio and IR visibilities depend on
baseline length.

In the case of LDN~1622 the radio-IR conversion factors given in
Table~\ref{table:corr} do show some variation at 12~$\mu$m when
comparing cases A and B.  In order to approximately account for the
$\theta = 12~$arcmin chop throw of \citet{fin02}, we restrict our
analysis to baselines in excess of $\theta^{-1}$~$\lambda$, or
286~$\lambda$, and use the non-differenced dataset. In this case,
$a_{100\mu\mathrm{m}}/10^{-3} = 0.92\pm 0.11$, which is
$\sim$2~$\sigma$ higher than the value of $0.70\pm0.01$ listed in
Table~\ref{table:corr} for the full dataset.

The tentative detection of spinning dust in LDN~1622 by \cite{fin02}
is based on a rise in flux density from 5~GHz to 9.75~GHz. Since their
5~GHz data were not chopped because of hardware limitations, the
rising SED could simply reflect the missing spatial frequencies.

We nonetheless confirm the tentative detection of a rising spectrum by
\cite{fin02}: after scaling units, the value for the dimensionless
$a_{100\mu\mathrm{m}}$ at 9.75~GHz is $1.5~10^{-4} \pm 0.5~10^{-4}$,
which by comparison with our value for $k > 286$~$\lambda$ implies a
spectral index $\alpha_{9.75~\mathrm{GHz}}^{31~\mathrm{GHz}} =
+1.57\pm0.31$.

\subsection{CBI spectral index}

Estimating a spectral index from the 10~CBI channels is difficult for
an extended object such as LDN~1622 because of varying
$uv-$coverage. Flux loss, due to missing spatial frequencies, is
greater in the high-frequency channels than in the low-frequency
channels.

\subsubsection{Estimates from MEM models}

Assuming that the MEM model is a good approximation to the sky signal,
a single spectral index $\alpha$ can be varied to minimize
\begin{eqnarray}
\chi^2 & = & \sum_i \| ^\mathrm{m}\!V_i(\alpha) -\, ^\mathrm{o}\!V_i \|
^2/\sigma_i^2  ~,   \\ 
^\mathrm{m}\!V_i(\alpha) &  = &
\left(\frac{\nu_i}{\nu_\circ}\right)^{\alpha} \,   {^\mathrm{m}}\!V_i(\nu = \nu_\circ), \label{eq:specindex}
\end{eqnarray}
where the sum extends over all baselines and all channels.
$\nu_\circ$ is the reference frequency used by MockCBI to scale the
intensity map by the input spectral index $\alpha = 0$ to the
frequency of the $i^\mathrm{th}$ visibility data point. Note that
although in this application MockCBI internally uses $\alpha = 0$, the
model visibilities $^\mathrm{m}\!V_i$ still bear a frequency
dependence through the $uv-$coverage and the primary beam. We optimize
$\chi^2$ by finding the root of $\partial \chi^2 / \partial \alpha$.

The entropy term does not depend on $\alpha$ since it is calculated on
the model image, which is kept constant for all channels in our
implementation. Yet the inclusion of a regularizing entropy biases the
spectral index estimates. In the case of LDN~1622, pure $\chi^2$
reconstructions, with $\lambda = 0$, result in noisy model images,
while in the absence of data a pure MEM reconstruction, with $\lambda
\rightarrow \infty$, defaults to a flat image, whose intensity is
$M/e$ (see Appendix~\ref{sec:memalgo}). Increasing values of $\lambda$
result in smoother model images, and the lower frequency channels
recover more flux from the model images than the higher frequency
channels.

We confirmed by simulation that $\alpha$ is recovered in pure $\chi^2$
reconstructions. The highest value of $\lambda$ for which the
resulting spectral index is not significantly biased is $\lambda =
1$. To obtain this limiting value we simulated the CBI observations on
template maps. We fit a model image and a single spectral index to
simulated visibilities on the processed {\em IRAS}~12$\mu$m and
~25$\mu$m templates. The CBI-simulated visibilities are calculated
with MockCBI using $\alpha = 0$. We ran our MEM algorithm 90 times,
with exactly the same settings as for the CBI models, feeding as input
the CBI-simulated visibilities with the addition of 90 different
realisations of Gaussian noise, as explained in Point~\ref{pt:noise}
of Appendix~\ref{sec:modval}. We rejected models that converged early on a
local minimum by requiring a minimum number of iterations
$N_\mathrm{iter}$.  For reference the $\lambda = 1$ CBI model
converged in $N_\mathrm{iter}= 29$ iterations. The average value of
best-fit indices in the simulations, without $N_\mathrm{iter}$ cutoff,
is $\alpha = 0.13 \pm 0.21$ for {\em IRAS}~12$\mu$m and $\alpha = 0.12
\pm 0.18$ for {\em IRAS}~25$\mu$m.  $\langle \alpha \rangle$ decreases
with increasing $N_\mathrm{iter}$, until it reaches the input value at
$N_\mathrm{iter} = 19$ for both {\em IRAS}~12$\mu$m and {\em
IRAS}~25$\mu$m. The resulting spectral index is $\alpha =0.009 \pm
0.172$ for {\em IRAS}~12$\mu$m and $\alpha =0.004 \pm 0.162$ for {\em
IRAS}~25$\mu$m, and satisfactorily close to zero, in the sense that
the systematic bias due to the smoothness introduced by the entropy
term is of order $+0.01$.

The 26--36~GHz CBI spectral index we obtained from the MEM modelling
with $\lambda = 1$ is $\alpha_\mathrm{CBI} = -0.38 \pm 0.13$.

\subsubsection{Estimates by cross-correlation with template maps}

Spectral indices are sometimes inferred by reference to a template
image, as in Sec.~\ref{sec:lowfreqindex}. The 31~GHz sky image of
LDN~1622 is assumed to follow exactly a template image, say {\em
IRAS}~100$\mu$m, so that the CBI image is a scaled version of the
reference image, and $V(\nu_i) = a_i V_\mathrm{template}$, where
$\{\nu_i\}_{i=1}^{10}$ are the CBI channel frequencies and where
$V_\mathrm{template}$ are CBI-simulated visibilities. The scaling
factor $a$ can be obtained as explained in Point~\ref{pt:acoef} of
Appendix~\ref{sec:modval}. The spectral behavior of the CBI
visibilities is thus cast into the scale coeficients.

However this strategy yields inconsistent results because it is
difficult to find an ideal reference image. Using the four {\em IRAS}
band, and averaging the 10 CBI channels in two frequencies, 28.5~GHz
and 33.5~GHz, we obtain spectral indices that depend strongly on the
reference template and on baseline range. For the full range of
baselines, $\alpha$ varies from $\alpha = -0.24 \pm 0.16$ for {\em
IRAS}~12$\mu$m to $\alpha = -0.06 \pm 0.15$ for {\em IRAS}~60$\mu$m,
the other {\em IRAS} bands giving intermediate values. For the
non-differenced dataset and $uv$-radii in excess of 120~$\lambda$, we
obtain values ranging from $\alpha = -1.12 \pm 0.30$ for {\em
IRAS}~12$\mu$m to $\alpha = -0.75 \pm 0.31$ for {\em
IRAS}~25$\mu$m. All of these alternative CBI-{\em IRAS}
cross-correlations could equally well be used to infer a spectral
index. But the difference between the extremal values obtained above
is greater than 3~$\sigma$, and is therefore significant. These
results are reported here to emphasize the systematic uncertainties
involved in determinations of spectral energy distributions inferred
by cross-correlations.
 
\subsection{Integrated SED} \label{sec:sed}

\begin{figure}
\epsscale{.5}  \plotone{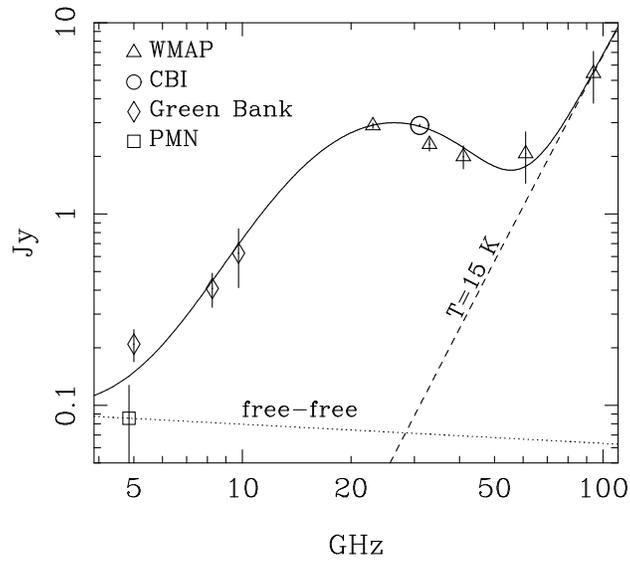}
\caption{SED of LDN~1622. The solid line is a fit to the data,
  composed of a free-free component, a modified blackbody at 15~K with
  a 1.7 emissivity index representative of traditional dust emission,
  and the \citet{dl98b} spinning dust emissivities.
  \label{fig:L1622_SED}}
\end{figure}

We extracted fluxes from the {\em WMAP}, {\em IRAS}, and PMN surveys
using a circular aperture with a diameter equal to the FWHM of the CBI
primary beam at 31~GHz, or 45~arcmin. In order to compare with the CBI
measurement, we also subtract a background level given by the flux
density in the CBI reference field (offset by 8~min to the East). For
all maps the reference field is essentially devoid of emission
compared to the object field. To take into account flux loss, the CBI
flux density we discuss here is that obtained by scaling the {\em
IRAS}~100$\mu$m flux density (see Sec.~\ref{sec:ground}). The existing
data on the integrated SED of LDN~1622 are summarised in
Fig.~\ref{fig:L1622_SED} and Table~\ref{table:L1622_SED}.


\begin{deluxetable}{lrr}
\tabletypesize{\small} \tablecaption{SED of LDN~1622.
\label{table:L1622_SED}} \tablewidth{0pt} \tablehead{  \colhead{$\nu$/GHz}
& \colhead{$F_\nu$/Jy} &  }
\startdata 
    4.85 & $  (8.6\pm 4.2)~10^{-2}  $ & PMN  \\  
    5.00 & $  (2.1\pm 0.4)~10^{-1}  $ & Green Bank$^a$ \\
    8.25 & $  (4.1\pm 0.8)~10^{-1}  $ & Green Bank$^a$ \\
    9.75 & $  (6.2\pm 2.1)~10^{-1}  $ & Green Bank$^a$ \\
    23.0 & $  2.9 \pm 0.09  $         & {\em WMAP} \\ 
    31.0 & $  2.9\pm 0.04  $          & CBI$^a$ \\
    33.0 & $  2.3\pm 0.18  $          & {\em WMAP} \\ 
    41.0 & $  2.0\pm 0.28  $          & {\em WMAP} \\ 
    61.0 & $  2.1\pm 0.62  $          & {\em WMAP} \\ 
    94.0 & $  5.4\pm 1.6   $          & {\em WMAP} \\ 
 3000    & $  (4.1\pm 0.41)~10^{3} $  & {\em IRAS} \\
 5000    & $  (1.1\pm 0.11)~10^{3} $  & {\em IRAS} \\  \enddata
\tablecomments{$^a$: measurements inferred by cross-correlation with
  {\em IRAS}~100~$\mu$m.}
\end{deluxetable}

The spectrum of the emissivity per unit proton column density in
LDN~1622 can be fit with the spinning dust emissivities of
\citet{dl98b}\footnote{available at
\tt{http://www.astro.princeton.edu/~draine/dust/dust.mwave.html}}, as
first shown by \citet{fin04}. The data points are fit with a mixture
of free-free emission, a modified blackbody representative of
traditional dust emission, and the spinning dust emissivities. We
require that a 15~K modified blackbody, with a 1.7 emissivity index,
crosses the 90~GHz {\em WMAP} point\footnote{Attempting to fit the
{\em IRAS}~100$\mu$m point and the {\em WMAP} W band simultaneously
resulted in excessively low emissivity indices, or in an unrealistic
sub-mm peak. The bulk of the dust in LDN~1622 is thus characterised by
at least 2 modified black bodies.}.  The spinning dust emissivities
depend on environment, and we confirm the result of \citet{fin04} that
the SED is best fit with a mixture of CNM and WNM emissivities
\citep[as defined by][]{dl98b}, with a fraction of 37$\pm$5\% CNM and
63$\pm$11\% WNM and a proton colummn averaged over the CBI primary
beam of $N_H = 1.24~10^{22}$~cm$^{-2}$ (somewhat less than
$2.4~10^{22}$~cm$^{-2}$, the value used by \citet{fin02} referring to
the peak extinction value).

The spectral indices obtained from the {\em WMAP} data are
$\alpha_{23}^{33} = -0.62 \pm 0.23$ and $\alpha_{33}^{41} = -0.68 \pm
0.73$, or $\alpha_{23}^{41} = -0.64 \pm 0.24$, which is within
1~$\sigma$ from $\alpha_\mathrm{CBI} = -0.38 \pm 0.13$. Combining all
measurements gives a 30~GHz index $\alpha_\mathrm{30~GHz} = -0.44 \pm
0.11$.

\clearpage
\clearpage



%
%
%


%

\section{Conclusions}  \label{sec:conc}

The CBI observations of LDN~1622 resulted in the first cm-wave
continuum image of a dark cloud, at frequencies where traditional
emission from dust is not expected. The CBI data follow a tight
correlation with the far-IR emission, confirming that the 31~GHz
emission is nonetheless related to dust.

Under visual inspection the 31~GHz map is closer to the {\em
IRAS}~12~$\mu$m and {\em IRAS}~25~$\mu$m maps than to the {\em
IRAS}~100~$\mu$m map. To quantify the IR--radio similarities we
calculate the cross-correlation of the 31~GHz images with each of the
{\em IRAS} images.  We find a trend for a decreasing cross-correlation
with wavelength, such that the 31~GHz--12~$\mu$m comparison has the
highest cross-correlation.


The mid-IR -- cm-wave correlation in LDN~1622 indicates that the
cm-wave continuum emission arises in a shell coincident with the PDR
at the surface of LDN~1622 exposed to the Ori~OB~1b UV field. The
closer match between the 31~GHz and 12~$\mu$m images can be
interpreted as support for spinning dust.  Alternatively the 31~GHz
continuum may stem from a mechanism of molecular continuum emission at
31~GHz or a dense molecular forest spread over 26--36~GHz.

We suspect the reason why the mid-IR -- cm-wave correlation was not
previously detected in other objects, or in the diffuse ISM, is
because the {\em IRAS}~12$\mu$m maps are contaminated by many more
point sources than the {\em IRAS}~100~$\mu$m maps. The stellar
emission at mid-IR wavelengths has no counterpart in cm-waves, as
shown here in the case of LDN~1622.

The 10 CBI channels allow estimating a spectral index
$\alpha_\mathrm{CBI} = -0.38 \pm 0.13$. Combining all measurements we
obtain $\alpha_\mathrm{30~GHz} = -0.44 \pm 0.11$.

\acknowledgments

This article benefitted from the constructive comments of an anonymous
referee that motivated Sec.~\ref{sec:sed}, the discussion on
unresolved radiosources, Fig.~\ref{fig:YSOSED} and its discussion in
Appendix~C.  S.C. acknowledges support from Fondecyt grant 1030805,
and from the Chilean Center for Astrophysics FONDAP 15010003.  We
gratefully acknowledge the generous support of Maxine and Ronald
Linde, Cecil and Sally Drinkward, Barbara and Stanely Rawn, Jr., Fred
Kavli, and Rochus Vogt.  This work is supported by the National
Science Foundation under grant AST 00-98734. We acknowledge the use of
NASA's {\em SkyView} facility ({\tt http://skyview.gsfc.nasa.gov})
located at NASA Goddard Space Flight Center.

\appendix

\section{MEM algorithm}  \label{sec:memalgo}

The MEM algorithm was programmed by us and fits model visibilities,
calculated on a model image, to the observed visibilities. The
free-parameters of our MEM model are the pixels in the model
$170\times170$ image, $\{I(x_i,y_i)\}_{i=1}^{170\times170}$ . We set
to zero all pixels that fall outside a region of the sky where the
expected noise is larger than a specified value. In practice, for one
pointing as is the case here, this means restricting the number of
free pixels to those that fall within a user-supplied radius from the
phase center.



The relatively small number of visibilities for the CBI ($\sim 1000$
for each on-off cycle) allows one to work in the $uv-$plane and fit
for the observed visibilities directly, rather than work in the sky
plane and deconvolve the synthesized beam. We did not apply any
gridding of the visibilities \citep{bri99}, which we postpone to a
future development of our code. The use of a direct fourier transform
in our current implementation is time consuming.

The model functional we minimize is $L = \chi^2 - \lambda S$, with
\begin{equation}
\chi^2 = \sum_i \| ^\mathrm{m}\!V_i -\, ^\mathrm{o}\!V_i \|
^2/\sigma_i^2  ~, \label{eq:chi2}
\end{equation}
where the symbol $\| z \|$ stands for the modulus of a complex number
$z$, the sum extends over all visibilities (i.e., the sum runs over
10 channels and 78 baselines), $\sigma_i$ is the root-mean-square
(rms) noise on the corresponding visibility, $^\mathrm{o}\!V_i$ stands
for the observed visibilities, and the model visibilities
$^\mathrm{m}\!V_i$ are given by
\begin{equation} ^\mathrm{m}\!V(u_i,v_i) = \int_{-\infty}^{+\infty} A_\nu(x,y)
I_\nu(x,y)\exp\left[-2\pi i (u_ix+v_iy)\right]
\frac{dx\,dy}{\sqrt{1-x^2-y^2}}  ~, \label{eq:vmodel}
\end{equation}
where $A_\nu(x,y)$ is the CBI primary beam and $x$ and $y$ are the
direction cosines relative to the phase center in two orthogonal
directions on the sky.  The model visibilities are calculated using
the MockCBI program (see below). We assume a flat spectral index for
the model image, i.e. $I_\nu = I(31~$GHz) over the 10 CBI channels.

We use the entropy $S= - \sum_i I_i \log ( I_i / M )$, where
$\{I_i\}_{i=1}^{N}$ is the model image and $M$ is a small intensity
value taken as the noise estimated by difmap and divided by 10000. We
also investigated an entropy term of the form $S_b = - \sum_i \log I_i
/ F$, where $F = \sum_i I_i $, obtaining essentially the same results.

Image positivity is enforced by clipping. All intensities below the
threshold value of $M$ are set equal to $M$. Our choice for the image
entropy is such that the entropy term minimizes the need for clipping
with a diverging derivative at zero intensities. However we caution
that the true sky signal in our differenced observations may not be
strictly positive: sources in the reference field act as negative
signal.

The entropy is used as a regularizing term. Because the reconstruction
is degenerate in the sense that we have more free parameters than data
points, pure $\chi^2$ reconstructions lead to artificially low values
of reduced $\chi^2$, so that $\chi^2$ models end up fitting the noise
(i.e. the residual image is artificially flat at the locus of free
parameters).  The parameter $\lambda$ was adjusted by hand and kept
fixed during the optimization. Intermediate values of $\lambda$ from
infinity to zero recover the sum of object signal and noise in
gradually increasing detail. The exact value of $\lambda$ is set by
trial and error, requiring that $\chi^2$ (Eq.~\ref{eq:chi2}) is close
to its expected value given by approximately twice the number of
imaginary data visibilities.  A dimensionless value of $\lambda = 5$
($\lambda = 5~10^{-9}$ for $S_b$) gave good results when
reconstructing on test images (see Fig.~\ref{fig:irasmem} below). We
obtain a reduced $\chi^2$ value of 1.04 for the CBI visibilities, with
25726 data points (i.e. twice the number of complex data
points). Reduced $\chi^2$ for the MEM models of the template {\em
IRAS} images is 0.99. The slightly larger $\chi^2$ for the CBI data is
probably due to faint sources in the reference field acting as
negative sky signal. The positivity requirement precludes modelling
such negative signal.


Convergence is achieved in $\sim 20$ iterations using the
Fletcher-Reeves conjugate-gradient algorithm from {\em Numerical
Recipes} \citep[NR]{pre96}, or $\sim 80$ if using the {\em Gnu
Scientific Library} (GSL, {\tt http://www.gnu.org/software/gsl/ }).
The GSL algorithm is double-precision, but is too slow for our needs
as it requires more gradient evaluations per iteration than in
NR. Thus the models presented in this work use the NR
implementation. One reconstruction takes about 30~min using the AMD
Athlon XP3000 processor, or 1h with an Intel Pentium 4 at 2.80~GHz.


%

\section{Model validation} \label{sec:modval}

To validate our MEM model we reconstructed the sky emission from model
visibilities, obtained by a simulation of CBI observations on
reference images (``CBI-simulated visibilities''). Simulation of the
CBI observations is performed with the MockCBI program (Pearson 2000,
private communication), which calculates the visibilities $V(u,v)$ on
the input images $I_\nu(x,y)$ with the same $uv$ sampling as a
reference visibility dataset (Eq.~\ref{eq:vmodel}). Thus MockCBI
creates the visibility dataset that would have been obtained had the
sky emission followed the template.

\begin{figure}
\epsscale{1.}
\plotone{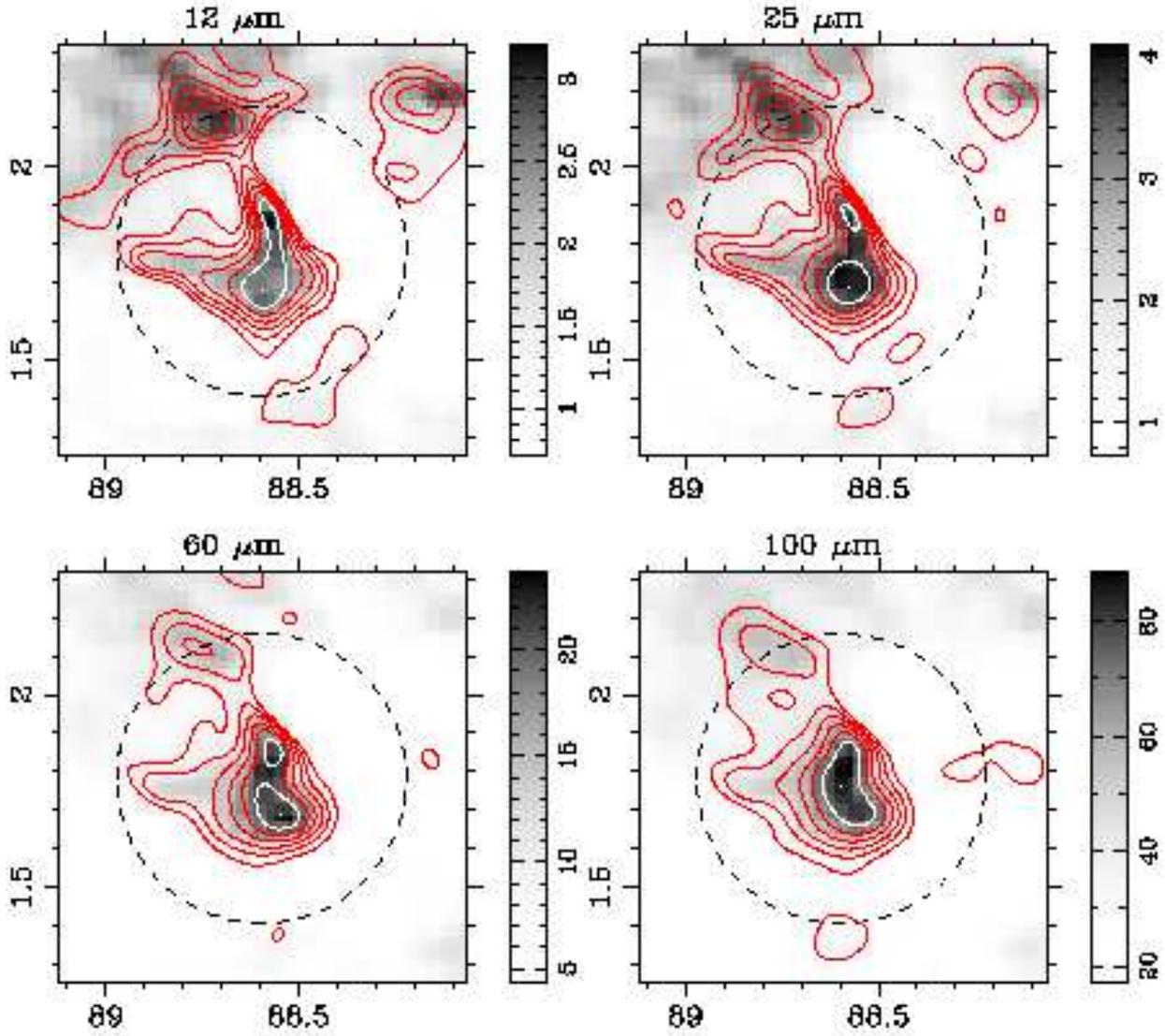}
\caption{Input {\em IRAS} template maps in grey scale, with overlays
  of our average MEM reconstructions from CBI-simulated data and 90
  different realisations of gaussian noise. The $x$ and $y$ axis are
  J2000 R.A. and Dec, in degrees of arc. Flux units are MJy~sr$^{-1}$.
  \label{fig:irasmem}}
\end{figure}

We used as reference images the maps of LDN~1622 in the four {\em
IRAS} bands, as downloaded from {\em SkyView} ({\tt
http://skyview.gsfc.nasa.gov}). The procedure is as follows:
\begin{enumerate}
\item Subtract conspicuous mid-IR point-sources in the 12~$\mu$m and
25~$\mu$m 3$\times$3~degree fields. We fit elliptical gaussians on a
second order polynomial surface. Only one of these point-sources
coincides with the object itself, namely L1622-10 (see
Appendix~\ref{sec:midirps}), but all contribute to the simulated
visibilities. L1622-10 is an entry in the {\em IRAS} Point Source
Catalog, so it was removed from the 12$\mu$m template by subtracting a
point source with L1622-10's tabulated 12$\mu$m flux of 1.027~Jy, with
a PSF given by the minimum width of the elliptical gaussian fits to
the other point sources (5.4~arcmin FWHM). We performed tests both
with and without subtraction of L1622-10.
\item Clip the {\em IRAS} images so that the minimum intensity value
is zero.  The processed images are shown on Fig.~\ref{fig:irasmem}.
\item Simulate CBI visibilities on the processed {\em IRAS} images
using MockCBI.
\item Cross-correlate the observed CBI visibilities with the model
visibilities to obtain 31~GHz--far-IR conversion factors, $a$:
$V(31~\mathrm{GHz}) = a V(IRAS)$, in the complex plane. We fit for $a$
by minimising $\chi^2 = \sum_i \| V(31~\mathrm{GHz}) - a V(IRAS) \|^2
/ \sigma_i^2 $, where the notation is the same as in
Eq.~\ref{eq:chi2}. \label{pt:acoef}
\item Divide the model visibilities by $a$ to obtain model
visibilities scaled to the 31~GHz values. Values for $a$ are given in
Table~\ref{table:corr}, for case A (differenced dataset).
\item Add gaussian noise to the complex model visibilities, (i.e. we
  assume the model visibilities have no noise), with a dispersion
  given by the root-mean-square (rms) noise on the corresponding CBI
  visibility.  \label{pt:noise}
\item Run the MEM reconstruction algorithm with the same parameters as
  for the observed CBI data.
\item Repeat the simulation 90~times with  90 different
  realisations of noise. 
\item Average the 90 model images. We tested that the measured scatter
  in the properties of the simulated reconstructions does not increase
  when increasing the number of noise realisations from 60 to 90
  (although 30 realisations was not enough).
\end{enumerate} 
We did not take into account the finite resolution of the {\em IRAS}
maps, which is due to the coarse pixelization used in the IRAS Sky
Survey Atlas maps available at {\em SkyView}. The net effect is that
the template resolution is lower than that of the CBI data. The mid-IR
point sources allow estimating that the natural-weight synthesized
beam is 20\% larger for the {\em IRAS} simulations that for the CBI
data.

Fig.~\ref{fig:irasmem} shows the average MEM models overlaid on the
input maps, and allows judging by inspection the level of detail that
can be recovered from the CBI visibilities of LDN~1622. In this case
we used the full $uv$-coverage of the CBI, as in Fig.~\ref{fig:cbi}.


\clearpage
\clearpage

\section{Mid-IR point sources} \label{sec:midirps}



Although LDN~1622 figures in lists of starless cores
\citep{lee01,par04} it harbors an entry of the {\em IRAS} Point Source
Catalog\footnote{Infrared Astronomical Satellite Catalogs, 1988. The
Point Source Catalog, version 2.0, NASA RP-1190}, IRAS~05517+0151,
whose presence can be inferred from the {\em IRAS}~12$\mu$m image in
Fig.~\ref{fig:irasmem}, because the peak of emission at a position of
$(88.58, 1.87)$ is bright and unresolved, and stands out over the
diffuse emission. IRAS~05517+0151 is coincident within the
uncertainties with an entry from the 2MASS catalog \citep{cut03},
2MASS 05542277+0152039, and with the binary pre-main-sequence star
L1622-10 \citep[J2000 RA: 05:54:26.8, Dec: +01:52:16,][]{rei93}.

\begin{figure}
\epsscale{.5}
\plotone{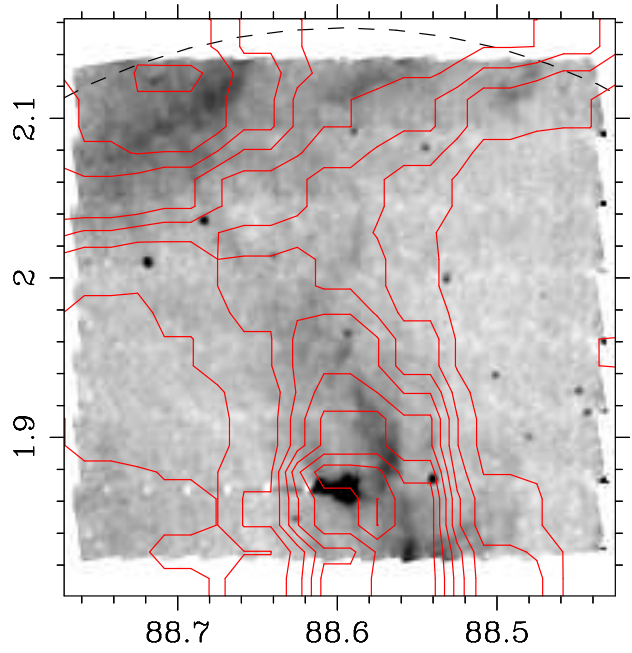}
\caption{The {\em IRAS} 12~$\mu$m contours overlaid on the 6.7~$\mu$m
ISOCAM mosaic of LDN~1622 in grey scale (arbitrary units),
highlighting the presence of a YSO at $(88.58, 1.87)$. The dashed arc
traces the FWHM of the CBI primary beam. The $x$ and $y$ axis are
J2000 R.A. and Dec, in degrees of arc.
\label{fig:isocam}}
\end{figure}

The YSO is very clear as a saturated pixel in the ISOCAM\footnote{ISO
is an ESA project with instruments funded by ESA Member States
(especially the PI countries: France, Germany, the Netherlands and the
United Kingdom) and with the participation of ISAS and NASA. The ISO
TDT and AOT codes for the image used here are 69802905 and C01, and
the observer is P.~Andr\'e.}  6.7~$\mu$m image of LDN~1622 presented
by \citet{bac00}. Note that LDN~1622 is curiously listed as LDN~1672
in \citet{bac00}, and the orientation of the image is not as that
obtained from the {\em ISO} archive, and is thus probably wrong. For
these reasons we present in Fig.~\ref{fig:isocam} an overlay of the
{\em IRAS}~12$\mu$m emission in contours on the upright ISOCAM
6.7~$\mu$m map. The 12~$\mu$m peak and the saturated region at
6.7~$\mu$m are coincident.
\clearpage
\clearpage

\begin{figure}
\epsscale{.5}
\plotone{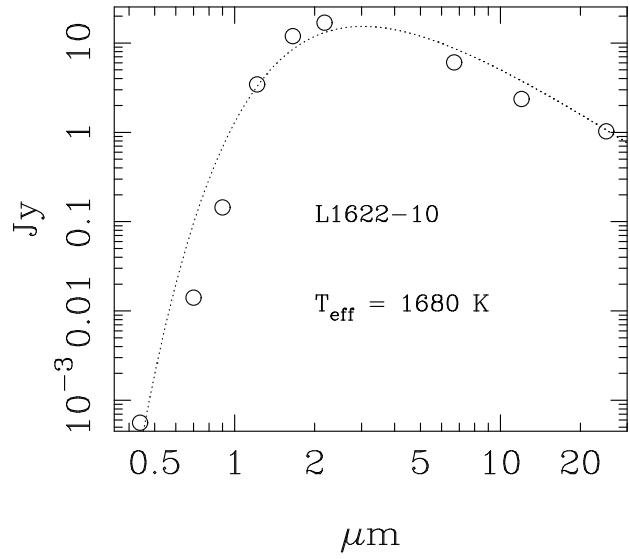}
\caption{The spectral energy distribution of the most conspicuous YSO
  coincident with LDN~1622. The dotted line is a blackbody fit to the
  data points, shown in circles.
\label{fig:YSOSED}}
\end{figure}

L1622-10 is probably a T-Tauri binary \citep{rei93}. We extracted the
B, R, and I, photometry of L1622-10 from the USNO-B10 Catalog
\citep{mon03}, and constructed the SED shown on
Fig.~\ref{fig:YSOSED}. An estimate of the ISO flux is included at
6.7~$\mu$m, but is assigned zero weights because the presence of pixel
glitches spreading away from L1622-10 suggests the detector may be
saturated. The {\em IRAS} 25~$\mu$m flux is also assigned zero weight
because of uncertainties in the nebular contamination.  A simple
blackbody fit gives a temperature of 1680$\pm$50~K, using conservative
uncertainties on the data points, and a linear size of 0.2$\pm$0.01~AU
for a distance of 120~pc. The integrated luminosity of the blackbody
fit is thus 2.8~L$_\odot$, corresponding to a 1.4~M$_\odot$
main-sequence star, which confirms that the YSO is a low-mass object.


\clearpage
\clearpage


\begin{thebibliography}{}


\bibitem[Bacmann et al.(2000)]{bac00} Bacmann, A., Andr\'e, P., Puget,
      	J.-L., Abergel, A., Bontemps, S., Ward-Thompson, D., 2000, \aap,
      	361, 555 
\bibitem[Banday et al.(2003)]{ban03} Banday, A.J., Dickinson, C.,
  Davies, R.D., Davis, R.J., G\'orski, K.M., 2003, \mnras, 345, 897
\bibitem[Becker et al.(1991)]{bec91} Becker, R. H., White, R. L., Edwards, A. L. 1991, \apjs, 75, 1
\bibitem[Bevington \& Robinson (1992)]{bev92} Bevington, P.R.,
  Robinson, D.K.,   1992, in ``Data Reduction and Error Analysis for
  the Physical Sciences'', 2$^\mathrm{nd}$ edition, McGraw-Hill
\bibitem[Boumis et al.(2001)]{bou01} Boumis, P., Dickinson, C.,
Meaburn, J., Goudis, C. D., Christopoulou, P. E., L\'opez, J. A., Bryce,
M., Redman, M. P., 2001, \mnras, 320, 61
\bibitem[Briggs et al.(1999)]{bri99} Briggs, A.S., Schwab, F.R.,
  Sramek, R.A., 1999, in ``Synthesis Imaging in Radio Astronomy II'',
  ASP Conference Series 180, 127
\bibitem[Casassus et al.(2004)]{cas04} Casassus,S., Readhead, A.C.S.,
  Pearson, T.J.,  Nyman, L.\AA., Shepherd, M.C., Bronfman, L., 2004,
  \apj, 603, 599
\bibitem[Condon et al.(1998)]{con98} Condon, J. J., Cotton, W. D.,
  Greisen, E. W., Yin, Q. F., Perley, R. A., Taylor, G. B., Broderick,
  J. J., 1998, \aj, 115, 1693
\bibitem[Condon el al.(1993)]{co93} Condon, J.J., Griffith, M.R. \& Wright, A.E., 1993, \aj, 106, 1095
\bibitem[Cutri et al.(2003)]{cut03} Cutri, R.M., et al., 2003,
     ``The 2MASS All-Sky Catalog of Point Sources'', University of
     Massachusetts and Infrared Processing and Analysis Center
     (IPAC/California Institute of Technology).
\bibitem[de\,Oliveira-Costa et al.(2002)]{deol02} de Oliveira-Costa, A,
Tegmark, M., Finkbeiner, D.P., Davies, R.D., Gutierrez, C.M., Haffner,
L.M., Jones, A.W., Lasenby, A.N., Rebolo, R., Reynolds, R.J., Tufte,
S.L., Watson, R.A., 2002, \apjl, 567, 363
\bibitem[de\,Oliveira-Costa et al.(1999)]{deol99} de Oliveira-Costa,
A., Tegmark, M., Guti\'errez, C.M., Jones, A.W., Davies, R.D., Lasenby,
A.N., Rebolo, R., Watson, R.A., 1999, \apjl, 527, 9
\bibitem[D\'esert et al.(1990)]{des90} D\'esert, F.-X., Boulanger, F.,
  Puget, J.L., 1990, \aap, 237, 215
\bibitem[Draine \& Lazarian(1998a)]{dl98a} Draine, B.T., Lazarian, A., 1998,
\apjl, 494, L19
\bibitem[Draine \& Lazarian(1998b)]{dl98b} Draine, B.T., Lazarian, A., 1998,
\apj, 508, 157
\bibitem[Draine \& Lazarian(1999)]{dl99} Draine, B.T., Lazarian, A., 1999,
\apj, 512, 740
\bibitem[Draine \& Li(2001)]{dra01} Draine, B.T., Li, A., 2001, \apj, 551, 807
\bibitem[Elmegreen(2002)]{elm02} Elmegreen, B. G., 2002, \apj, 564, 773
\bibitem[Finkbeiner (2004)]{fin04} Finkbeiner, D.P., 2004, \apj, 614, 186, 
\bibitem[Finkbeiner et al.(1999)]{fin99} Finkbeiner, D.P., Davis, M.,
Schlegel, D.J., 1999, \apj, 524, 867
\bibitem[Finkbeiner et al.(2002)]{fin02} Finkbeiner, D.P., Schlegel, D.J.,
Frank, C., Heiles, C., 2002, \apj, 566, 898
\bibitem[Gaustad et al.(2001)]{gau01} Gaustad, J.E., McCullough, P.R.,
Rosing, W., Van Buren, D., 2001, \pasp, 113, 1326
\bibitem[Gautier et al.(1992)]{gau92} Gautier, T.N. III, Boulanger,
  F., P\'erault, M., Puget, J.L., 1992, \aj, 103, 1313
\bibitem[Heiles et al.(2000)]{hei00} Heiles, C., Haffner, L.M.,
Reynolds, R.J., Tufte, S.L., 2000, \apj, 536, 335
\bibitem[Li \& Draine(2001)]{li01} Li, A., Draine, B.T., 2001,\apj, 554, 778
\bibitem[Lagache(2003)]{lag03} Lagache, G.,2003, \aap, 405, 813

\bibitem[Laureijs et al.(1989)]{laur89} Laureijs, R.J., Chlewicki, G.,
Wesselius, P.R., Clark, F. O., 1989, \aap, 220, 226
\bibitem[Lee et al.(2001)]{lee01} Lee, C.W., Myers, P. C., Tafalla, M., 2001, \apjs, 136, 703
\bibitem[Lee \& Myers(1999)]{lee99} Lee, C.W., Myers, P. C., 1999,
  \apjs, 123, 233
\bibitem[Leitch et al.(1997)]{lei97} Leitch, E.M., Readhead, A.C.S., Pearson,
  T.J., Myers, S.T., 1997, \apjl 486, L23
\bibitem[Lynds(1962)]{lyn62} Lynds, B.T., 1962, \apjs, 7, 1
\bibitem[Maddalena et al.(1986)]{mad86} Maddalena, R.J., Morris, M.,
Moscowitz, Thaddeus, P., 1986, \apj, 303, 375
\bibitem[Monet et al.(2003)]{mon03} Monet D.G., Levine S.E.,
Casian B., et al.,  2003, \aj, 125, 984.
\bibitem[Padin et al.(2002)]{pad02} Padin, S., et al, 2002, \pasp, 114, 83
\bibitem[Page et al.(2003)]{pag03} Page, L., Nolta, M. R., Barnes, C.,
Bennett, C. L., Halpern, M., Hinshaw, G., Jarosik, N., Kogut, A.,
Limon, M., Meyer, S. S., Peiris, H. V., Spergel, D. N., Tucker, G. S.,
Wollack, E., Wright, E. L., 2003, \apjs, 148, 39.
\bibitem[Park et al.(2004)]{par04} Park, Y.-S., Lee, C. W., Myers,
  P. C., 2004,   \apjs, 152, 81
\bibitem[Pearson et al.(2003)]{pea03} Pearson, T. J., Mason, B. S.,
	Readhead, A. C. S., Shepherd, M. C., Sievers, J. L.,
	Udomprasert, P. S., Cartwright, J. K., Farmer, A. J., Padin,
	S., Myers, S. T., Bond, J. R., Contaldi, C. R., Pen, U.-L.,
	Prunet, S., Pogosyan, D., Carlstrom, J. E., Kovac, J., Leitch,
	E. M., Pryke, C., Halverson, N. W., Holzapfel, W. L.,
	Altamirano, P., Bronfman, L., Casassus, S., May, J., Joy, M.,
	2003, \apj, 591, 556
\bibitem[Press et al.(1996)]{pre96} Press, W.H., Flannery, B.P.,
  Teukilsky, S.A., Vettering, W., Y., 1996, Numerical Recipes
  (Cambridge, Cambridge Univerisity Press)
\bibitem[Reipurth \& Zinnecker(1993)]{rei93} Reipurth, B., Zinnecker,
  H., 1993, \aap, 278, 81
\bibitem[Shepherd(1997)]{she97} Shepherd, M.C., 1997, in Astronomical
  Data Analysis Software and Systems VI, ed. G~Hunt \& H.E.~Payne, ASP
  conference series, v125, 77-84 ``Difmap: an interactive program for
  synthesis imaging''.
\bibitem[Van Dishoeck(2004)]{vand04} Van Dishoeck, E.F., 2004, \araa,
  42, 119
\bibitem[Watson et al.(2005)]{wat05} Watson, R. A., Rebolo, R.,
Rubi\~no-Mart\'in, J. A., Hildebrandt, S., Guti\'errez, C. M.,
Fern\'andez-Cerezo, S., Hoyland, R. J., Battistelli, E. S., 2005,
\apj, 624, 89
\bibitem[Wheelock et al.(1991)]{whe91} Wheelock, S., et al., 1991,
IRAS Sky Survey Atlas Explanatory Supplement
\bibitem[Wilson et al.(2005)]{wil05} Wilson, B.A., Dame, T.M.,
  Masheder, M.R., Thaddeus, P., 2005, \aap, 430, 523
\bibitem[Wright(1998)]{wri98} Wright, E.L., 1998, \apj, 496, 1



\end{thebibliography}
\end{document}